\documentclass[12pt]{spieman}  
\usepackage{amsmath,amsfonts,amssymb}
\usepackage{graphicx}
\usepackage{setspace}
\usepackage{tocloft}
\usepackage{algorithm}
\usepackage{algpseudocode}
\usepackage{comment}
\usepackage{xcolor}
\usepackage{color, soul}

\title{CT Reconstruction using Diffusion Posterior Sampling conditioned on a Nonlinear Measurement Model}

\author[a,c]{Shudong Li}
\author[a]{Xiao Jiang}
\author[b]{Matthew Tivnan}
\author[d]{Grace J. Gang}
\author[c,*]{Yuan Shen} 
\author[a,*]{J. Webster Stayman} 
\affil[a]{Department of Biomedical Engineering, Johns Hopkins University, Baltimore, MD, USA, 21218}
\affil[b]{Department of Radiology, Harvard Medical School and Massachusetts General Hospital, Boston, MA, USA, 02114}
\affil[c]{Department of Electronic Engineering, Tsinghua University, Beijing, China, 100084}
\affil[d]{Department of Radiology, University of Pennsylvania, Philadelphia, PA, USA, 19104}

\cftpagenumbersoff{figure}
\cftpagenumbersoff{table} 
\begin{document} 
\maketitle

\begin{abstract}
Diffusion models have been demonstrated as powerful deep learning tools for image generation in CT reconstruction and restoration. Recently, diffusion posterior sampling, where a score-based diffusion prior is combined with a likelihood model, has been used to produce high quality CT images given low-quality measurements. This technique is attractive since it permits a one-time, unsupervised training of a CT prior; which can then be incorporated with an arbitrary data model. However, current methods rely on a \textit{linear} model of x-ray CT physics to reconstruct or restore images. While it is common to linearize the transmission tomography reconstruction problem, this is an approximation to the true and inherently nonlinear forward model. We propose a new method that solves the inverse problem of \textit{nonlinear} CT image reconstruction via diffusion posterior sampling. We implement a traditional unconditional diffusion model by training a prior score function estimator, and apply Bayes rule to combine this prior with a measurement likelihood score function derived from the nonlinear physical model to arrive at a posterior score function that can be used to sample the reverse-time diffusion process. This plug-and-play method allows incorporation of a diffusion-based prior with generalized nonlinear CT image reconstruction into multiple CT system designs with different forward models, without the need for any additional training. We develop the algorithm that performs this reconstruction, including an ordered-subsets variant for accelerated processing and demonstrate the technique in both fully sampled low dose data and sparse-view geometries using a single unsupervised training of the prior.
\end{abstract}

\keywords{CT reconstruction, deep learning reconstruction, diffusion model, diffusion posterior sampling, deep learning}

{\noindent \footnotesize Corresponding authors: \textbf{*}J. Webster Stayman,  \linkable{web.stayman@jhu.edu} }
{\noindent \footnotesize\textbf{*}Yuan Shen,  \linkable{shenyuan\_ee@tsinghua.edu.cn} }

\begin{spacing}{2}   
\section{Introduction}
\label{sec:intro}  
Neural networks have been widely investigated for CT reconstruction and have shown impressive performance capabilities \cite{kang2017deep, wu2017iterative, jing2022training}. Recently, diffusion models have been shown to be particularly powerful deep learning tools for CT reconstruction and restoration \cite{kazerouni2023diffusion, chung2022improving ,muller2022diffusion, liu2022dolce, xia2022patch}. Such methods have generally been based on denoising diffusion probabilistic models (DDPM) \cite{ho2020denoising} and score-based generative diffusion models through stochastic differential equations (SDEs) \cite{song2020score}. 
Most diffusion models for CT reconstruction have been trained in a supervised manner, where the conditional data input is known in both training and generation \cite{liu2022dolce, xia2022patch, tivnan2023fourier}. Recent research on image denoising and restoration has demonstrated the potential of unsupervised training by leveraging the power of posterior sampling, where the conditional data input is only available during the reverse-time diffusion process \cite{song2021solving, chung2022diffusion, chung2022improving}. The unsupervised training does not assume a fixed measurement process during training, and can thus be flexibly incorporated with different measurement models without retraining. In these efforts, measurement model for diffusion posterior sampling has been presumed to be linear. Lopez-Montes et al.\cite{lopezmontes2023diffusion} have recently applied such an approach using a linearized model to tomographic data.

Nevertheless, it is well-known that the transmission tomography forward model that relates the underlying attenuation values to the measurements is nonlinear. While the model is commonly linearized for mathematical convenience, such approximation will lose the opportunity to precisely describe those more sophisticated imaging system, thus can result in deviation from the true underlying data model and can cause problems in many circumstances including models for very low dose data \cite{bushe2023unbiased}, beam-hardening \cite{hsieh2000iterative}, and detector blur \cite{tilley2015model}. In this work, we propose an approach for CT image reduction using diffusion posterior sampling that is conditioned on a nonlinear measurement model (which we will refer to as {\em DPS Nonlinear}). Specifically, we seek to solve the inverse problem of CT image reconstruction via DPS with a score-based diffusion prior and a nonlinear physics-driven measurement likelihood term. This method guides the updates of the reverse-time diffusion process so that the final results are consistent with the raw (nonlinear) measurements from a CT imaging system. Our approach trains a traditional score-based diffusion model for unconditional CT image generation and applies Bayes rule by adding the gradient of the log-likelihood, i.e., from the non-linear physical model to achieve the posterior score function necessary for DPS. The method is plug-and-play, meaning it does not require any extra training for application to a different CT system measurement model. {A preliminary development and evaluation of these proposed methods was presented in Li et al.\cite{li2024diffusion} Significant changes from that presentation include training on a more realistic anatomy, adoption of a Poisson noise model, and more in-depth evaluations including comparisons to related deep learning approaches.}


\section{Method}
\subsection{Diffusion Posterior Sampling conditioned on a Nonlinear Physical Model}
\begin{figure}[h]
\centering
   \centering
   \begin{tabular}{c}
   \includegraphics[height=5.5cm]{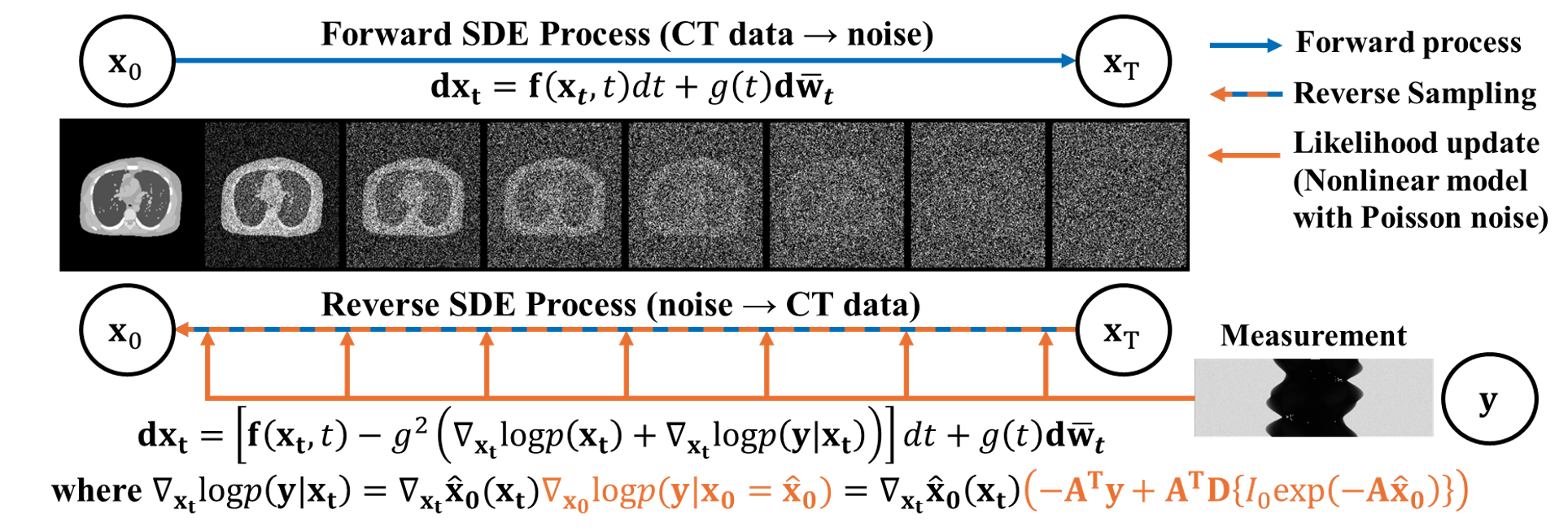}
   \end{tabular}
   \caption[main workflow] 
   {\label{fig:main_workflow}{The workflow of the proposed DPS Nonlinear method. The blue arrow represents a forward stochastic process defined by the equation in the top wherein $\mathbf{x_0}$, a ``clean'' ground truth CT image is successively degraded in intermediate noisy images, $\mathbf{x_t}$, indexed by time $t$, to an image of Gaussian noise, $\mathbf{x_T}$,. A reverse stochastic process is defined at the bottom that is conditioned on measurement data, $\mathbf{y}$. The orange arrow represents this conditioning and is based on a general nonlinear physical measurement model.}}
\end{figure}

{Figure \ref{fig:main_workflow} shows the general workflow of our proposed DPS Nonlinear method, which combines diffusion posterior sampling with innovative likelihood updates based on a nonlinear physical measurement model (indicated in orange). } This approach requires estimation of the prior distribution of CT images using a score-based generative model \cite{ho2020denoising, song2020score}. Specifically, a diffusion model is adopted wherein a forward stochastic process {(top portion of Figure~\ref{fig:main_workflow})} is defined where an image is perturbed by successively added noise. A reverse process {(Figure~\ref{fig:main_workflow}, bottom)} is then defined where we may sample images from the estimated prior distribution starting from a noise image. 
Specifically, we follow prior work \cite{ho2020denoising, song2020score} where the forward diffusion process is governed by
\begin{equation}
\label{eq:forward}
    \mathbf{x_t} = \sqrt{\bar{\alpha}_t}\mathbf{x_0} + {\sqrt{(1 - \bar{\alpha}_t)}}\mathbf{\epsilon_{noise}} \;\;\; t\in(0,T],
\end{equation}
\noindent where $\mathbf{\epsilon_{noise}} \sim \mathcal{N}(0,I)$ is Gaussian noise, $\mathbf{x_0} \sim \mathbf{p_{data}}$ denote a true image sampled from a training dataset, $\bar{\alpha}_t$ is a variance scheduler such that the final noise is close to a predefined noise distribution $\mathbf{\pi(x)}$, and the process is defined on $t\in(0,T]$. The final sample $\mathbf{p_{t=T}}$ usually follows the predefined noise distribution $\mathbf{\pi(x)}$ and it does not contain any information of training dataset $\mathbf{p_{data}}$ or $\mathbf{p_{t=0}}$.  To recover the data in a tractable distribution, which can be achieved by reversing the perturbation forward process in Equation~(\ref{eq:forward}) to get the reverse-time SDE given by 
\begin{equation}
\label{eq:reverse}
    \mathbf{dx_t} = [\mathbf{f}(\mathbf{x_t}, t) + g^2(t)\nabla_{\mathbf{x_t}}\mathrm{log}p(\mathbf{x_t})]dt + g(t)\mathbf{d\bar{w}_t} \;\;\; t\in(0,T],
\end{equation}
\noindent where $\mathbf{f}$ is drift coefficient term for $\mathbf{x_t}$ defined by $\mathbf{f}(\mathbf{x_t}, t) = -1/2 \beta_t\mathbf{x_t}$, $g$ is the diffusion coefficient for $\mathbf{x_t}$ defined by $g(t)=\sqrt{\beta_t}$, $\beta_t = -1/\sqrt{\bar{\alpha}_t}\frac{d}{dt}\bar{\alpha}_t$, $\bar{\mathbf{w}}$ is a standard Wiener process when reverse-time flows from T to 0, and $dt$ is an infinitesimal negative timestep. The continuous structure also allows for the production of data samples from $\mathbf{p_0}(\mathbf{x_0}|\mathbf{y})$ if the posterior $\mathbf{y}$ and $\mathbf{p_t}(\mathbf{y}|\mathbf{x_t})$ is known. Equation~(\ref{eq:reverse}) then can be transformed into a conditional reverse-time SDE:
\begin{equation}
\label{eq:conditioned_reverse}
    \mathbf{dx_t} = [\mathbf{f}(\mathbf{x_t}, t) + g^2(t)\nabla_{\mathbf{x_t}}\mathrm{log} p(\mathbf{x_t}|\mathbf{y})]dt + g(t)\mathbf{d\bar{w}_t} \;\;\; t\in(0,T].
\end{equation}
\noindent Furthermore, applying the Bayes rule, the conditioned reverse stochastic process in Equation~(\ref{eq:conditioned_reverse}) for diffusion posterior sampling is given by 
\begin{equation}
    \label{eq:reverse_posterior}
    \mathbf{d\mathbf{x_t}} = [\mathbf{f}(\mathbf{x_t}, t) + g^2(t)(\nabla_{\mathbf{x_t}}\mathrm{log} p(\mathbf{x_t}) + \nabla_{\mathbf{x_t}}\mathrm{log} p(\mathbf{y}|\mathbf{x_t}))]dt + g(t)\mathbf{d\bar{w}_t} \;\;\; t\in(0,T].
\end{equation}
\noindent Following previous score matching methods\cite{song2020score, hyvarinen2005estimation}, the prior score $\nabla_{\mathbf{x_t}}\mathrm{log} p(\mathbf{x_t})$ is estimated by a neural network called score model $\mathbf{s_\theta}(\mathbf{x_t},t)$ using the following loss function:
\begin{equation}
    \label{eq:score_model}
    \theta^*=\arg \min_\theta\  E_{t \sim u(\epsilon,1), \mathbf{x_t} \sim p(\mathbf{x_t} | \mathbf{x_0}), \mathbf{x_0} \sim \mathbf{p_{data}}}[||\mathbf{s_\theta}(\mathbf{x_t},t) - \nabla_{\mathbf{x_t}}\mathrm{log} p(\mathbf{x_t}|\mathbf{x_0})||^2_2],
\end{equation}
\noindent where the network parameters, $\theta$, are estimated through a training process and the expected loss is evaluated over the uniform distribution timestep $t\in(0,T]$ with degraded images $\mathbf{x_t}$. The degraded images are generated in the forward process in Equation (\ref{eq:forward}) by perturbing ground truth images $\mathbf{x_0}$ from the training dataset for the corresponding timestep $t$. Plugging the score estimate into Equation~(\ref{eq:reverse_posterior}) yields the resulting reverse-time SDE:
\begin{equation}
    \label{eq:reverse_sampling}
    \mathbf{dx_t} = [\mathbf{f}(\mathbf{x_t}, t) + g^2(t)(\mathbf{s_{\theta^*}}(\mathbf{x_t},t)+ \nabla_{\mathbf{x_t}}\mathrm{log} p(\mathbf{y}|\mathbf{x_t}))]dt + g(t)\mathbf{d\bar{w}_t} \;\;\; t\in(0,T].
\end{equation}
\noindent Following Chung et al.\cite{chung2022diffusion}, the gradient of the likelihood term, $\nabla_{\mathbf{x_t}}\mathrm{log} p(\mathbf{y}|\mathbf{x_t})$, can be approximated as follows:
\begin{equation}
    \label{eq:likelihood}
\nabla_{\mathbf{x_t}}\mathrm{log} p(\mathbf{y}|\mathbf{x_t})=\nabla_{\mathbf{x_t}}\mathbf{\hat{x}_0}(\mathbf{x_t})\nabla_{\mathbf{x_0}}\mathrm{log} p(\mathbf{y}|\mathbf{x_0} = \mathbf{\mathbf{\hat{x}_0}})
    \; \text{where} \; \mathbf{\hat{x}_0} = \frac{1}{\sqrt{\bar{\alpha}_t}}(\mathbf{x_t}+(1-\bar{\alpha}_t)\mathbf{s_{\theta^*}}(\mathbf{x_t},t)),
\end{equation}
\noindent where $\mathbf{\hat{x}_0}$ denotes an approximation of $\mathbf{x_0}$ from $\mathbf{x_t}$, and $\nabla_{\mathbf{x_t}}$ denotes the gradient with respect to $\mathbf{x_t}$. For $\nabla_{\mathbf{x_0}}\mathrm{log} p(\mathbf{y}|\mathbf{x_0} = \mathbf{\hat{x}_0})$, we propose to apply a nonlinear physical model defined by a {Poisson} distribution:
\begin{equation}
    \label{eq:physical_model}
{\mathbf{y} \sim Poisson\{\mathbf{\bar{y}}\}
\;\;\; \text{where} \;\;\;    
    \mathbf{\bar{y}}(\mathbf{x_0}) = I_0\exp(-\mathbf{A}\mathbf{x_0}),}
\end{equation}
\noindent where $\mathbf{\bar{\mathbf{y}}}$ denotes the mean measurement vector, $\mathbf{y}$ denotes the real projection coming from different CT system, $\mathbf{x_0}$ denotes the image of attenuation coefficients in transmission tomography, $\mathbf{A}$ is the linear projection matrix defined for a given CT system geometry. {$I_0$ is a scalar defining the mean number of photons/detector pixel in bare beam. Based on Erdogan et al.{\cite{erdogan1999ordered}}, we may write:}
\begin{equation}
\label{eq:detail_explain_gradient_of_poisson}
{\nabla_{\mathbf{x_0}}\mathrm{log} p(\mathbf{y}|\mathbf{{x}_0}) = -\mathbf{A}^T\mathbf{y} + \mathbf{A}^TD\{I_0\exp(-\mathbf{A}\mathbf{{x}_0})\}}.
\end{equation}
\noindent where $\{\cdot\}^\text{T}$ denotes a matrix transpose with $\mathbf{A}^\text{T}$ representing backprojection. Equations~(\ref{eq:likelihood}) and Equations~(\ref{eq:detail_explain_gradient_of_poisson}) are plugged into Equation Equations~(\ref{eq:reverse_sampling}) to enable nonlinear physical model based diffusion posterior sampling in a reverse-time diffusion process.

\subsection{Training of Score Model in Forward Diffusion}

\noindent {CT Lymph Nodes Dataset \cite{roth2014new}} was used for training of score model. A total of {156} subjects were divided into training, validation, and test sets with a ratio of approximately {7:2:1} and then individual slices of {512 $\times$ 512} voxels were then extracted for each dataset. We adopted a schedule defined by $\bar{\alpha}_t = 1e^{-5t}$ with $t\in(0,T]$ where $T=1$. For the score model, we adopt a U-Net \cite{ronneberger2015u} with an attention residual module and time embedding module\cite{song2020score}. For training, we applied Adam optimizer, setting the batch size to 8 and the initial learning rate to 0.001. We trained our score model on training set over 1200 Epochs with early stopping of {100 Epochs}. This trained score model was subsequently used for CT reconstruction for different CT system models.

\subsection{Reconstruction using Diffusion Posterior Sampling}

\noindent For the reverse-time diffusion, we plug in the likelihood score function and generate the samples with the prior score function from the trained score model. The algorithm includes a time-dependent coefficient scheduler $\lambda_t$, which balances the weighting between the likelihood score function and the prior score function over the process of reverse-time diffusion. {We designed a heuristic weighting scheduler $\lambda_t$ that can greatly stabilize the reverse process in our experiments:
\begin{equation}
\label{eq:lambda}
\lambda_t = \frac{k}{||\nabla_{\mathbf{x_t}}\log p(\mathbf{y}|\mathbf{x_t})||^2_2},
\end{equation}
\noindent where weighting coefficient $k$ is a scalar that is weighted by the magnitude of the gradient likelihood and that can be tuned manually under different scenarios.} Following Chung et al. \cite{chung2022diffusion}, taking the derivative of the prior score function $\mathbf{s_{\theta^*}}(\mathbf{x_t},t)$ with respect to the input $\mathbf{x_t}$ is accomplished by calculating the Jacobian matrix using backward propagation in the score model. We adopted the standard Variance Preserving Stochastic Stochastic Differential Equations (VP-SDEs) to discretize the reverse process into $N$ bins. The final DPS Nonlinear reconstruction algorithm is summarized in the following {Algorithm \ref{alg:DPS Nonlinear Poisson}}. 
\noindent
\begin{algorithm}[ht]
\begin{algorithmic}[1]
\caption{DPS Nonlinear}
\label{alg:DPS Nonlinear Poisson}
\Require $\mathbf{x_1} \sim \mathcal{N}(0,I)$, \ {$\mathbf{\hat{y}} \gets I_0\exp(-\mathbf{A}\mathbf{x_y})$, \ $\mathbf{y} \sim Poisson(\mathbf{\hat{y}})$}, \ 
$\lambda(t)$,\
$\Delta t \gets \frac{1}{N}$\;
    \State \# Diffusion Posterior Sampling
    \For{\texttt{$i = N-1$ to $0$}}:
        \State $t \gets (i+1) / N$
        \State $\mathbf{\hat{s}} \gets \mathbf{s_\theta^*}(\mathbf{x_t},t)$ 
        \State $\mathbf{\hat{x}_0} \gets \frac{1}{\sqrt{\bar{\alpha}_t}}(\mathbf{x_t} + (1 - \bar{\alpha}_t)\mathbf{\hat{s}})$
        \State {$\mathbf{\Delta_{x_t}} \gets - \mathbf{A}^T\mathbf{y} + \mathbf{A}^TD\{I_0\exp(-\mathbf{A}\mathbf{{x}_0})\}$} 
        \State $\mathbf{\Delta_{x_0}} \gets \frac{1}{\sqrt{\bar{\alpha}_t}}(1 + (1 - \bar{\alpha}_t)\nabla_{\mathbf{x_t}}\mathbf{\hat{s}}(\mathbf{x_t}))$
        \State \# $\nabla_{\mathbf{x_t}}\mathbf{\hat{s}}(\mathbf{x_t})$ is taking the backpropagation w.r.t $\mathbf{x_t}$
        \State $\mathbf{dx} \gets [\mathbf{f}(t)\mathbf{x_t} + g^2(t)(\mathbf{\hat{s}} + \lambda(t) \mathbf{\Delta_{x_0}} \mathbf{\Delta_{x_t}})]dt$
        \State $\mathbf{z} \sim \mathcal{N}(0,I)$
        \State $\mathbf{x_{t-\Delta t}} \gets \mathbf{x_t} - \mathbf{dx} + g(t)\sqrt{\Delta t}\mathbf{z}$
    \EndFor
\end{algorithmic}
\end{algorithm}

The probabilistic framework of the DPS model allows a separation of the prior model and the measurement model. This separation is explicit both mathematically in the application of Bayes rule, but also in the resulting algorithm which alternates between reverse diffusion updates and gradient-based updates based on the likelihood model. Thus, the DPS approach can be viewed as a way to merge model-based iterative reconstruction (MBIR) and a diffusion-based prior. Indeed, if the diffusion prior were removed, one would have a maximum-likelihood approach that uses a gradient ascent algorithm for iterations. This also suggests that one may leverage some of the many acceleration strategies that have been applied in MBIR to improve the efficiency of DPS. This was also recognized by Wenjun et al \cite{xia2023diffusion}. Here, we modify the DPS Nonlinear approach to use ordered-subsets.\cite{erdogan1999ordered, ahn2003globally, hudson1994accelerated}

In short, in classic MBIR, it was found that the gradient of the likelihood can be well-approximated by a subset of the projection data (usually a fraction of projection angles from regularly spaced data), which requires only a fraction of the computations associated with the full projection ($\mathbf{A}$) or backprojection ($\mathbf{A}^T$). To apply the same technique to DPS Nonlinear, we divide the projections into assigned number of subsets $N_{os}$. Instead of using the whole projection $\mathbf{y}$, updates are based on rotating subsets of projections, applied in turn $\mathbf{y_j}$ where $j\in\;\{1,2,3, ..., N_{os}\}$. One can interpret this with as Order-Subset nonlinear diffusion posterior sampling (\textit{OS-DPS Nonlinear}), given by
\begin{equation}
    \label{eq:os_reverse_posterior}
    \mathbf{d}\mathbf{x_t} = [\mathbf{f}(\mathbf{x_t}, t) + g^2(t)(\nabla_{\mathbf{x_t}}\mathrm{log} p(\mathbf{x_t}) + \nabla_{\mathbf{x_t}}\mathrm{log} p_j(\mathbf{y_j}|\mathbf{x_t})]dt + g(t)\mathbf{d}\mathbf{\bar{w}_t}\;\;\;t\in(0,T]
\end{equation}
\noindent where $j\in\{1, 2, 3, ..., N_{os}\}$ is the current index of subset, $t\in(0,T]$ is the timesteps. The subset $j$ can be calculated by $mod(i,  N_{os})$, which is taking the remainder of current index of iteration $i$ divided by number of total order subsets. The likelihood term for each order subset $\nabla_{\mathbf{x_t}}\mathrm{log} p_j(\mathbf{y_j}|\mathbf{x_t})$ is updated through the following equation
\begin{equation}
    \label{eq:os_posterior_sampling_physical_poisson}
    {
    \nabla_{\mathbf{x_0}}\mathrm{log} p_j(\mathbf{y_j}|\mathbf{x_0} ) = -\mathbf{A_j}^T\mathbf{y_j} + \mathbf{A_j}^T\mathbf{D}\{I_0\exp(-\mathbf{A_j}\mathbf{{x}_0})\}.
    }
\end{equation}
\noindent where $\mathbf{A_j}$ denotes subset projection matrices, and $\mathbf{A^T_j}$ are the associated subset backprojection matrices. The DPS Nonlinear method can be treated as a special case of OS-DPS Nonlinear when the $N_{os} = 1$. The OS-DPS Nonlinear algorithm is summarized in {Algorithm \ref{alg:OS-DPS Nonlinear Poisson}}.
\begin{algorithm}[h]
\label{alg:OS-DPS Nonlinear Poisson}
\caption{OS-DPS Nonlinear}
\begin{algorithmic}[1]
\Require $\mathbf{x_1} \sim \mathcal{N}(0,I)$, $\Delta t \gets \frac{1}{N}$, $\lambda(t)$
\Require {$\mathbf{\hat{y}_j} \gets I_0\exp(-\mathbf{A_j}\mathbf{x_y})$,\ 
$\mathbf{y_j} \sim Poisson(\mathbf{\hat{y}_j})$},\
$j \in \{1,2,3,...,N_{os}\}$\;
\For{$i = N-1 \ \text{to} \  0$}
    \State $t \gets (i+1)/N$    
    \State $j \gets mod(i+1, N_{os})$    
    \State $\mathbf{\hat{s}} \gets \mathbf{s_\theta^*}(\mathbf{x_t},t)$     
    \State $\mathbf{\hat{x}_0} \gets \frac{1}{\sqrt{\bar{\alpha}_t}}(\mathbf{x_t} + (1 - \bar{\alpha}_t)\mathbf{\hat{s}})$    
    \State {$\mathbf{\Delta_{x_t}} \gets -\mathbf{A_j}^T\mathbf{y_j} + \mathbf{A_j}^T\mathbf{D}\{I_0\exp(-\mathbf{A_j}\mathbf{{x}_0})\}$}  
    \State $\mathbf{\Delta_{x_0}} \gets \frac{1}{\sqrt{\bar{\alpha}_t}}(1 + (1 - \bar{\alpha}_t)\nabla_{\mathbf{x_t}}\mathbf{\hat{s}}(\mathbf{x_t}))$   
    \State \# $\nabla_{\mathbf{x_t}}\mathbf{\hat{s}}(\mathbf{x_t})$ is taking the backpropagation w.r.t $\mathbf{x_t}$    
    \State $\mathbf{dx} \gets [\mathbf{f}(t)\mathbf{x_t} + g^2(t)(\mathbf{\hat{s}} + \lambda(t) \mathbf{\Delta_{x_0}} \mathbf{\Delta_{x_t}}]dt$
    \State $\mathbf{z} \sim \mathcal{N}(0,I)$   
    \State $\mathbf{x_{t-\Delta t}} \gets \mathbf{x_t} - \mathbf{dx} + g(t)\sqrt{\Delta t}\mathbf{z}$
\EndFor
\end{algorithmic}
\end{algorithm}

\subsection{Evaluation Studies}

\noindent We evaluate DPS Nonlinear in {three} scenarios: 1) Fully sampled low-dose CT reconstruction; 2) Sparse-view CT reconstruction; {and 3) CT reconstruction across varying noise levels. These three scenarios share the same CT system geometry} with a source-to-detector distance of 1500~mm, and a source-to-axis distance of 800~mm. Projections are one-dimensional with 1024 - 1.556~mm pixels. For the fully sampled low-dose CT reconstruction, $I_0 = 10^3$ photons/pixel with 360 projections covering $360^\circ$. For the Sparse view CT reconstruction, $I_0 = 10^5$ photons/pixel with 30 projections cover $360^\circ$. {For CT reconstruction across multiple noise levels. We consider the fully sampled CT systems with varied incident fluence, where $I_0$ equals to $10^2$, $10^3$, $10^4$, and $10^5$ photons/pixel with 360 projections covering $360^\circ$.}

For {the fully sampled low-dose and sparse-view CT reconstructions}, we compute traditional filtered-backprojection (FBP) and model-based iterative reconstructions (MBIR) {(with and without total variation regularization \cite{vogel1996iterative}). Additionally, we compare DPS Nonlinear to two related diffusion approaches to investigate potential advantages. In particular, we seek to study the impact of the unsupervised model training and the explicitly nonlinear model of the proposed approach. Toward that end we select: 1) A diffusion technique that uses supervised training and conditioning on measurement data - DOLCE {\cite{xia2023diffusion}} and, 2) Diffusion posterior sampling based on a linearized measurement model {\cite{chung2022diffusion}} (which we will refer to as DPS Linear). The linearized forward measurement is defined as follows: 
\begin{equation}
\label{eq:Linearize measurement}
\mathbf{l} = \mathbf{A}\mathbf{x}, \;\;\; \text{where} \;\;\; \mathbf{l} = -\log(\mathbf{y}/I_0),
\end{equation}
where DPS Linear aims to solve the inverse of the linear reconstruction problem using diffusion posterior sampling. We further report the inference runtime of all three diffusion methods in the fully sampled low-dose scenario} 
Additional evaluations of OS-DPS Nonlinear were conducted using the fully sampled low-dose CT reconstruction scenario in a limited computation case. 

For each approach we report Peak Signal-to-Noise ratio (PSNR) and Structured Similarity (SSIM). {For the evaluation of diffusion posterior sampling based methods like DPS Linear and DPS Nonlinear, we propose an additional generation evaluation method: For the same noisy measurement, we run the posterior sampling to generate multiple results and calculate the bias image $\mathbf{x}_{bias}$ and standard deviation image $\mathbf{x}_{std}$ defined as follows:
\begin{equation}
\label{eq:Bias}
\mathbf{x}_{bias} = |\mathbb{E}[\mathbf{\hat{x}}] - \mathbf{x_0}|
\end{equation}
\begin{equation}
\label{eq:variance}
\mathbf{x}_{std} = \sqrt{\mathbb{E}[(\mathbf{\hat{x}} - \mathbb{E}[\mathbf{\hat{x}}])^2]}
\end{equation}
\noindent where $\hat{x}$ is all the generated sample images, $\mathbf{x_0}$ is the ground truth. We can summarize the bias and standard deviation by taking the root mean square of bias image $\mathbf{x}_{bias}$ and mean of standard deviation image $\mathbf{x}_{std}$, respectively}. All the methods and systems are implemented using PyTorch and LEAP projectors / backprojectors{\cite{kim2023differentiable}} {on the platform of NVIDIA GeForce RTX 3090Ti GPU with Intel(R) Xeon(R) CPU E5-2697 v4.} Detailed settings are defined in the following experiments.

\subsubsection{Low-dose CT Reconstruction Comparisons}

\noindent For MBIR, we adopt {$10000$ iterations using Adam gradient optimizer with initial step size set to $0.01$ with total variance regularization (TV). This large number of iterations and relatively small step size was used to ensure a highly converged image estimate. The TV weighting was selected as $6\times10^{-6}$ based on a subjective image quality assessment.
DOLCE is a supervised diffusion-based method which requires a FBP or MBIR volume as an input in addition to projection data. We implemented DOLCE using the same training and validation sets as DPS Nonlinear with a fully sampled low-dose FBP as an additional input. A low-dose FBP was also used as a conditional input for subsequent evaluations. For the DPS Linear diffusion model, we adopted the same model weights as DPS Nonlinear since only the conditional reverse process is different.} {A standard DDPM sampler \cite{ho2020denoising}} with $1000$ reverse sampling steps was applied for all three diffusion-based methods (DPS Nonlinear, DPS Linear, DOLCE). {For fair comparison between DPS Linear and DPS Nonlinear, we evaluated 16 different weighting coefficients $k$ in equation (10) from $70$ to $490$ with an interval of 30, and computed the root mean squared bias and standard deviation. (Based on the comparison of bias and standard deviation, we selected $k = 310$ as the best weighting for image comparison.)}

\subsubsection{Sparse view CT reconstruction Comparisons}

\noindent {We apply MBIR to the low-dose scenario with same parameters as the fully sampled scenario, except that the TV was chosen empirically to be $2\times10^{-5}$.}
{To evaluate the ability of a supervised diffusion approach to operate in a different imaging scenario than it was trained, we evaluate the performance of DOLCE (trained on the fully sampled scenario) applied to the sparse-view scenario.  Sparse-view FBP is used as a conditional input for DOLCE. Again, for DPS Linear and DPS Nonlinear, different $k$ coefficients were evaluated from 200 to 1320 with an interval of 80. (We found $k = 840$ based on the consideration of both root mean squared bias and standard deviation value.)}

\subsubsection{Reconstruction across a range of fluence/noise levels}

{
We evaluate DPS Nonlinear across a range of fluence levels. We optimize the weighting coefficient individually for each fluence level and find optimal values at $k=310$ for $I_0=10^2$, $k=780$ for $I_0=10^3$, $k=1080$ for $I_0=10^4$, and $k=1280$ for $I_0=10^5$ based on RMSB. We generate $400$ DPS samples and showcase the one closest to the mean PSNR (over those samples). We also compute the error map for that sample, and a standard deviation map showing variability over all samples. To further illustrate variability and susceptibility to hallucination as a function of fluence level, we present the variation in both global PSNR (across the entire image) and local PSNR (within a lung region) as a function of fluence. Moreover, we compute error images for the best and worst case samples (based on local PSNR).}

\subsubsection{OS-DPS Nonlinear Comparisons}

\noindent For evaluation of OS-DPS Nonlinear, we consider 1, {3}, 6, 12, {18}, 24, 36, 48, 60, {72, 120, and 180 ordered subsets} (OS) in the fully sampled low-dose scenario for the comparison. {The weighting coefficient is set to $k=150$ according to the same optimization strategy as in fully low-dose and sparse-view scenarios.} 
In addition to PSNR and SSIM, {we also plot out the bias and standard deviation images across all the ordered subsets.}

\section{Results}
\label{sec:sections}

\subsection{Low-dose CT Reconstruction}
\begin{figure}[h]
\centering
\begin{minipage}{\textwidth}
   \centering
   \begin{tabular}{c} 
   \includegraphics[height=7.6cm]{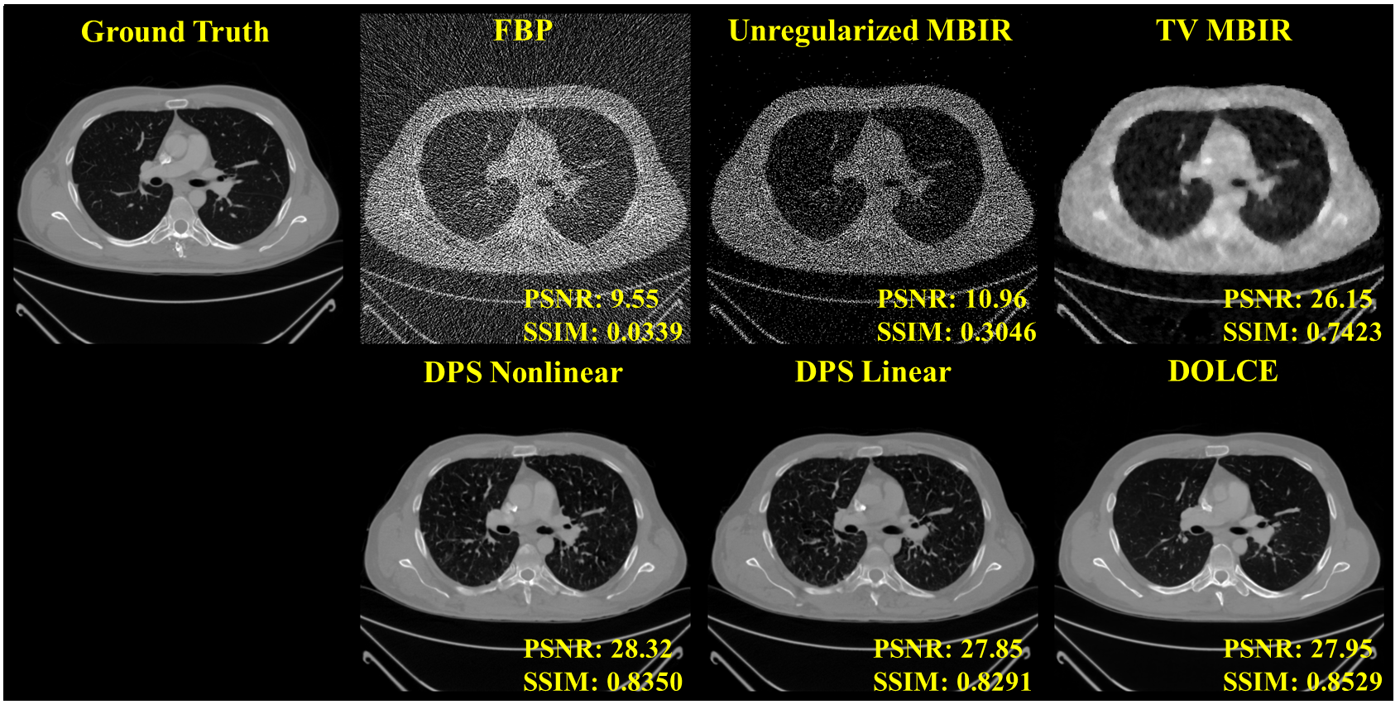}
   \end{tabular}
   \caption[Low dose image results] 
   { \label{fig:low_dose} {
Low-dose CT reconstructions comparing FBP, MBIR (without and with TV regularization), DPS Linear, DOLCE, and the proposed DPS Nonlinear approach. All images share a common colormap with W/L = 2000/0 HU.}}
   \end{minipage}
\end{figure}

{Figure {\ref{fig:low_dose}} shows the image results from the different reconstruction methods for the low dose CT scenario. The three diffusion-based methods yield sharp images with clear anatomical structures.} In contrast, the FBP reconstruction has heavy streaking associated with the very low x-ray fluence (particularly along the long axis and the spine where photon starvation effects are more pronounced.)  The MBIR approach {without regularization} is very noisy  {while total variance regularized version has reduced noise,} blurring obscures many of the fine details. The PSNR of our proposed technique was {$28.32$ dB, which is $0.67$ dB ($\sim2\%$) higher than DPS Linear and  $0.38$ dB ($\sim1\%$) higher than DOLCE. For SSIM, DOLCE achieves the best with $0.8529$. Our proposed DPS Nonlinear achieved $0.8350$, which is $0.0179$ lower than DOLCE ($\sim2\%$), but higher than DPS Linear. While one might expect DOLCE to outperform the other DPS techniques because of the supervised training where system-specific ground truth and degraded images are provided, we see that the relative performance among these diffusion methods is relatively close. We note that all diffusion methods show a degree of false feature generation (particularly in the fine detail structure of the lungs).}

\begin{figure}[h]
\centering
\begin{minipage}{\textwidth}
   \centering
   \begin{tabular}{c} 
   \includegraphics[height=4.7cm]{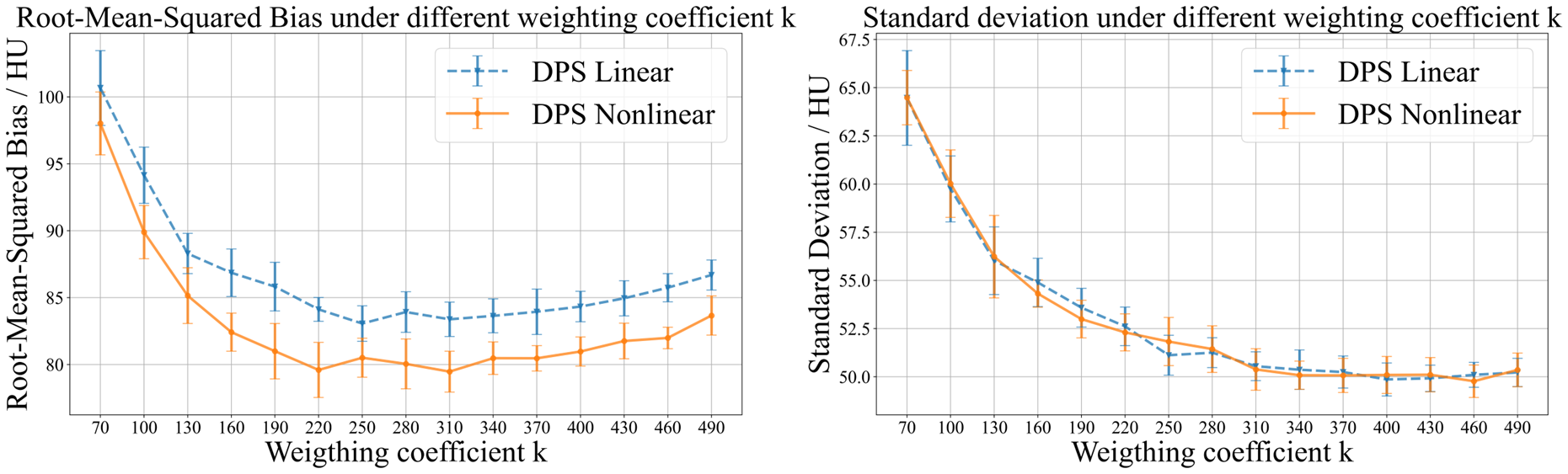}
   \end{tabular}
   \caption[Low dose DPS plot] 
   { \label{fig:DPS_Compare_under_low_dose}
{Bias and standard deviation comparison for low-dose CT reconstructions comparing the proposed DPS Nonlinear and DPS Linear approach. Root-mean-squared bias and standard deviation are reported in Hounsfield units (HU) based on 16 generated samples from diffusion posterior sampling for each weighting coefficient $k$ value. Error bars ($\pm~1.0$ standard deviation) are shown based on the variability over samples. The best weighting coefficient $k$ was $310$ based on the lowest root-mean-squared bias as well as the lowest standard deviation.}}
   \end{minipage}
\end{figure}

\begin{figure}[h]
\centering
\begin{minipage}{\textwidth}
   \begin{center}
   \begin{tabular}{c} 
   \includegraphics[height=3.6cm]{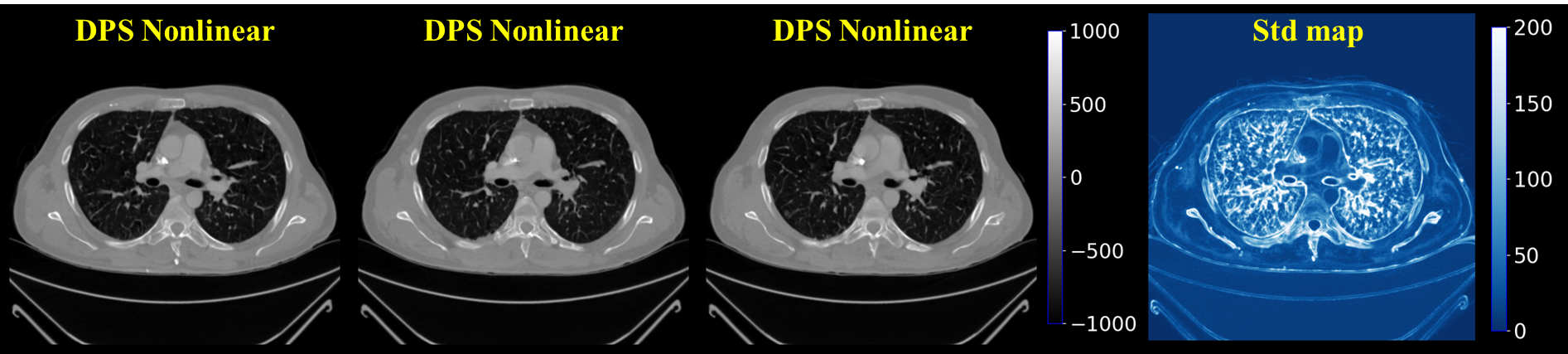}
   \end{tabular}
   \end{center}
   \caption[Low dose DPS results]
   {\label{fig:low_dose_posterior_sampling} 
Additional DPS sample reconstructions and standard deviation image from the same noisy measurement illustrating variability in the outputs. Colormap: W/L = 2000/0 HU for sample image and W/L = 500/250 HU for standard deviation image.}
\end{minipage}
   \end{figure}

{Figure \ref{fig:DPS_Compare_under_low_dose} shows the evaluation of DPS Linear and DPS Nonlinear under different weighting coefficients, $k$ values, in the low-dose scenario. We can see that DPS Nonlinear and DPS Linear share a similar performance curve line for root-mean-squared bias and standard deviation. For the best weighting coefficient, $k=310$, the proposed DPS Nonlinear achieves $\sim 4.7\%$ lower bias compared to DPS Linear while maintaining the similar standard deviation. Such biases have been noted in linearized MBIR reconstructions at low exposures due to the approximations to the Poisson model, and that same kind of bias appears for DPS Linear versus DPS Nonlinear (where no approximation is made).}
Figure \ref{fig:low_dose_posterior_sampling} further explores some of the generative aspects of the proposed DPS approach. Among several sample outputs applied to the same data, one can see both common features and variable features in the reconstruction. In general, DPS Nonlinear is more stable in the generation of low frequency general structures than the high frequency details. In particular, the fine details in the lung tend to be most variable with more consistency in major high contrast structures. This is particularly evident from a variance image computed from {16} of these DPS examples. Note the increased variability at tissue boundaries and in the smaller features in the lungs.

\subsection{Sparse-view CT Reconstruction}

\begin{figure}[h]
   \begin{center}
   \begin{tabular}{c} 
   \includegraphics[height=7.6cm]{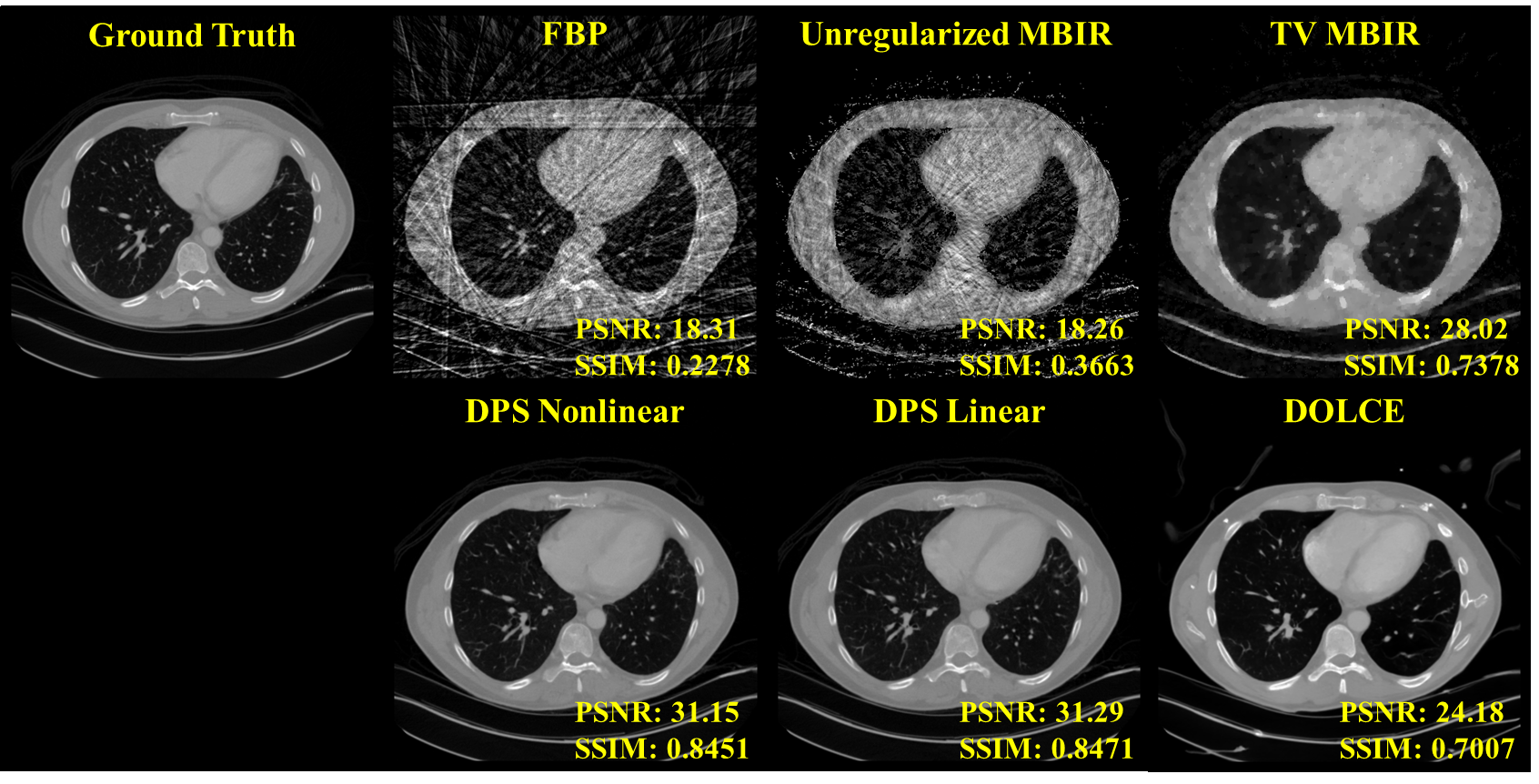}
   \end{tabular}
   \end{center}
   \caption[Sparse view image results] 
   { \label{fig:sparse_view} 
Sparse-view CT reconstructions comparing FBP, {MBIR (without and with TV regularization), DPS Linear, DOLCE, and the} proposed DPS Nonlinear approach. All images share a common colormap with W/L = 2000/0 HU.}
   \end{figure}

\noindent Figure \ref{fig:sparse_view} illustrates the flexibility of the DPS Nonlinear approach with application to a sparse-view CT reconstruction. Again, we see that the DPS Nonlinear {and DPS Linear} yields an image with sharp edges and recognizable features. In this case, there is better agreement with the underlying ground truth (suggesting, perhaps, less reliance on the generative nature of DPS and a better conditioned reconstruction problem). The FBP reconstruction exhibits obvious streaking due to undersampling, and MBIR shows significant blur and noise. {DOLCE (which was not trained for the sparse-view scenario) tends to have smoother features within the lungs and exhibits some significant false generative features including additional bones in the chest wall. In contrast, DPS Nonlinear (and Linear) appears to be much more robust in handling different measurement models}. In terms of PSNR, {DPS Nonlinear yielded $31.15$ dB versus DPS Linear with $31.29$ dB, and DOLCE with $24.18$ dB.} For SSIM, {DPS Nonlinear was $0.8451$ while DPS Linear and DOLCE was $0.8471$ and $0.7007$}, respectively. These quantitative measures agree well with perceived image quality.
   
\begin{figure}[h]
\centering
\begin{minipage}{\textwidth}
   \centering
   \begin{tabular}{c} 
   \includegraphics[height=4.8cm]{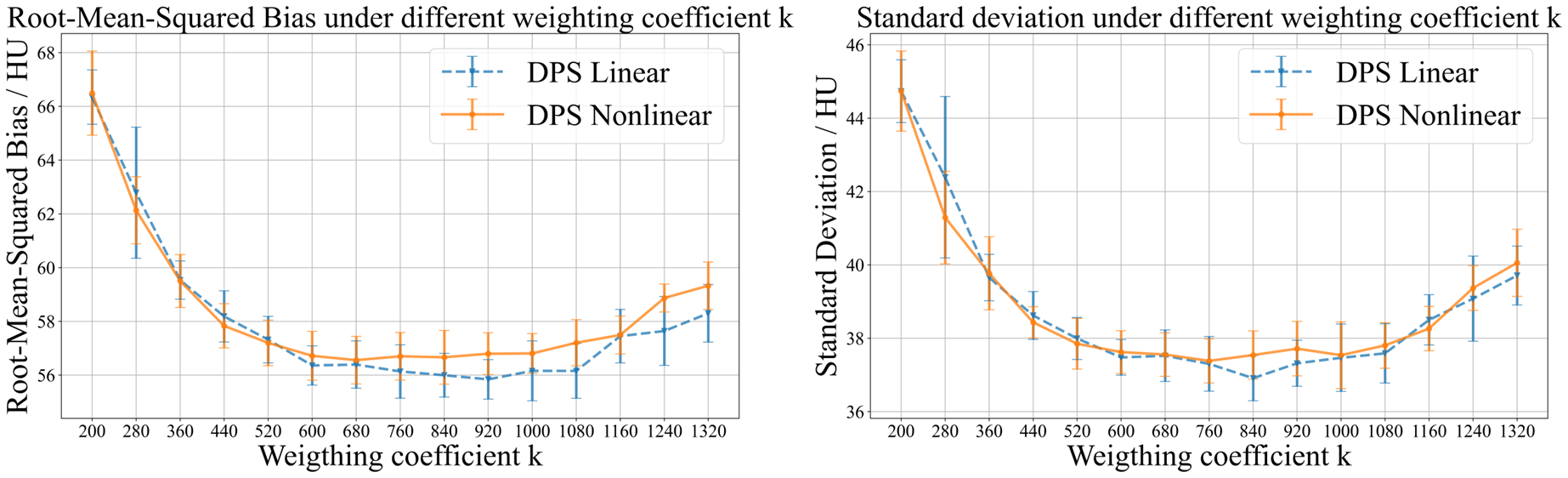}
   \end{tabular}
   \caption[Sparse view DPS plot] 
   { \label{fig:DPS_Compare_under_sparse_view}
{Bias and standard deviation comparison for sparse-view CT reconstructions comparing the proposed DPS Nonlinear and DPS Linear approach. Root-mean-squared bias and standard deviation are reported in Hounsfield units (HU) based on 16 generated samples from diffusion posterior sampling for each weighting coefficient $k$ value. Error bars ($\pm~1.0$ standard deviation) are shown based on the variability over samples. The best weighting coefficient $k$ was selected to be $840$.}}
\end{minipage}
\end{figure}

\begin{figure}[h]
\centering
\begin{minipage}{\textwidth}
   \begin{center}
   \begin{tabular}{c} 
   \includegraphics[height=3.6cm]{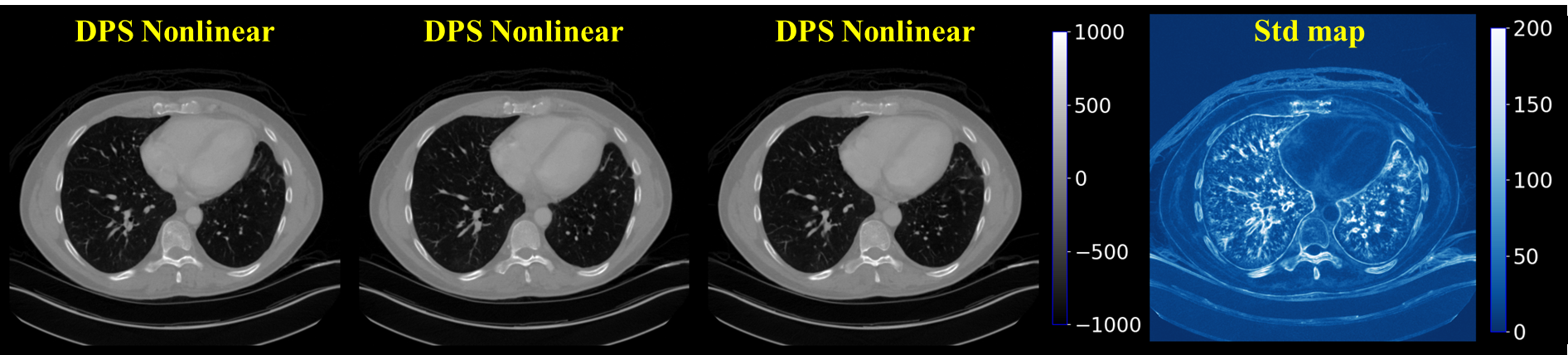}
   \end{tabular}
   \end{center}
   \caption[Sparse view DPS results]
   {\label{fig:sparse_view_posterior_sampling} 
Additional DPS sample reconstructions from the same data illustrating variability in the outputs. Colormap: W/L = 2000/0 HU for sample image and W/L = 500/250 HU for standard deviation image.}
\end{minipage}
\end{figure}

{Figure \ref{fig:DPS_Compare_under_sparse_view} shows the root-mean-squared bias and standard deviation of DPS Nonlinear and DPS Linear for the sparse-view scenario. Again, there are similar performance curves, where both approaches achieve nearly the same bias and standard deviation optima. Balancing noise and bias, we chose an optimal value of $k=840$. We do observe a slight difference in mean bias of around $1.2\%$ for DPS Linear. However, we note that this improvement is within the $\pm~1.0$ standard deviation error bars (based on variability over DPS samples), suggesting that this difference is not particularly significant.}
As in the low-dose situation, we generated {16} samples and computed a variance image. Three DPS Nonlinear reconstructions are shown to illustrate some of the variability associated with the stochastic reverse process. Three examples and the variance image are shown in Figure \ref{fig:sparse_view_posterior_sampling}. Similar to the low-dose scenario, DPS Nonlinear tends to be more stable in the generation of low frequency general structures than the high frequency details in the lung and at tissue boundaries.
   
\subsection{Reconstruction across a range of fluence/noise levels}

\begin{figure}[h]
\centering
\begin{minipage}{\textwidth}
   \begin{center}
   \begin{tabular}{c} 
   \includegraphics[height=10cm]{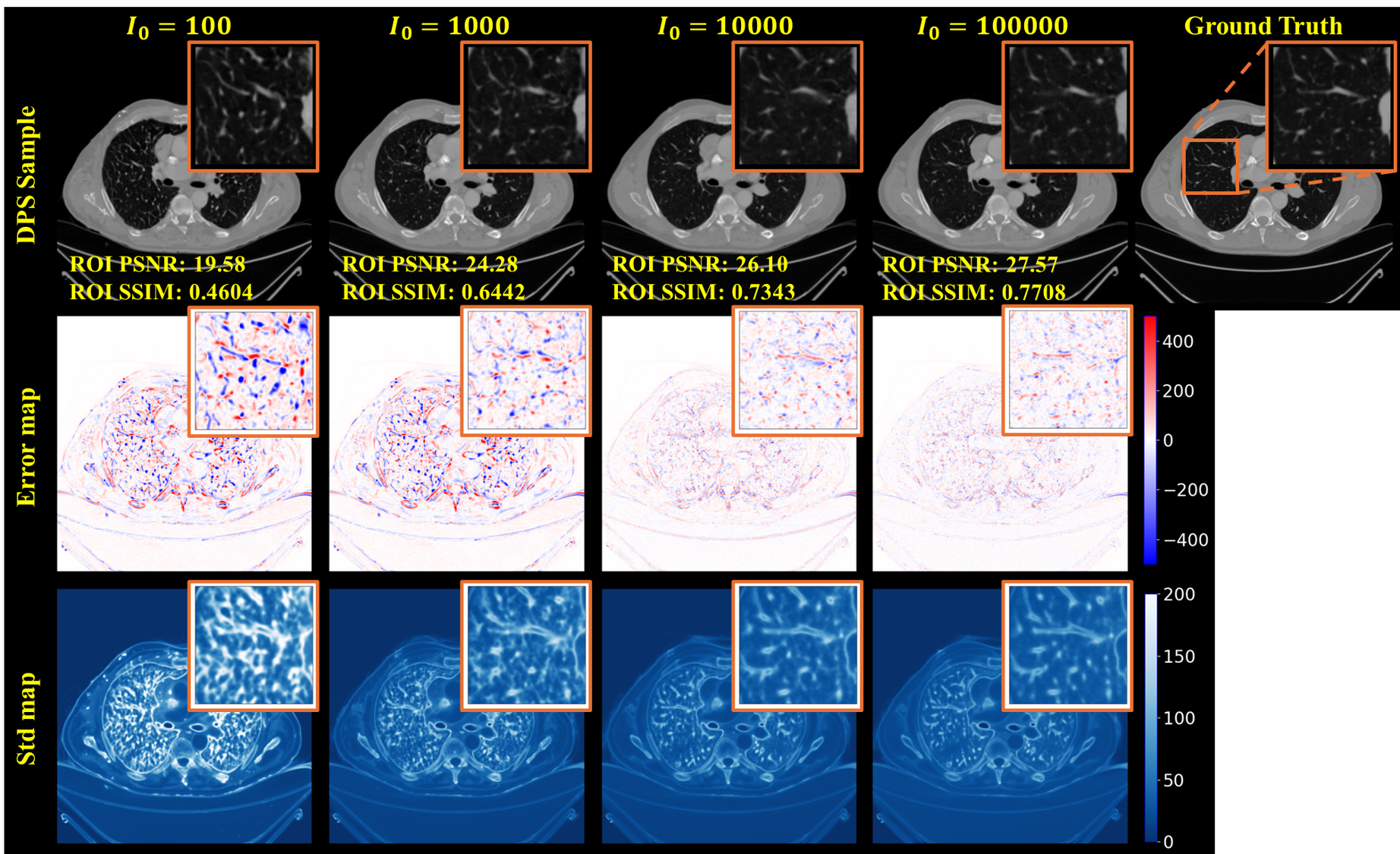}
   \end{tabular}
   \end{center}
   \caption[Multiple noise levels DPS results]
   {\label{fig:multiple_noise_realization}{ 
   DPS Nonlinear reconstructions from the measurement data with varying fluence levels from $I_0 = 100$ (noisiest) to $I_0 = 10^5$ (least noise). The top row shows reconstructed DPS samples that has closest PSNR to the mean PSNR of all $400$ generated samples. The second row shows the error map w.r.t the ground truth. The third row is the standard deviation map calculated based on $400$ samples. Colormap: W/L = 2000/0 HU for DPS Samples, W/L = 1000/0 HU for error map, W/L = 200/100 HU for std map.}}
\end{minipage}
\end{figure}

{DPS Nonlinear CT reconstruction results under different fluence levels are shown in Figure \ref{fig:multiple_noise_realization}. Among 400~DPS samples, we highlight the reconstruction that most closely achieves the mean PSNR in a local lung region of interest (ROI). As $I_0$ decreases, the mean ROI PSNR decreases from $27.57$ to $19.58$ dB. While the images appear sharp at the lowest exposures, one can see that the errors in the image are highly structured - appearing like typical lung features. Such hallucinations are more pronounced at the lowest exposure level. At the two highest exposures, errors are more consistent with the ground truth - appearing as differences in contrast (e.g. the feature is there but at a slightly different attenuation level). Moreover, the errors tend to be higher spatial frequency at higher fluence. Spatial variability is presented as a standard deviation map, where we see that the highest variability in samples occurs at features edges and in the texture of the lung, and is most pronounced at the lowest exposure.} 

\begin{figure}[h]
\centering
\begin{minipage}{\textwidth}
   \begin{center}
   \begin{tabular}{c} 
   \includegraphics[height=4.8cm]{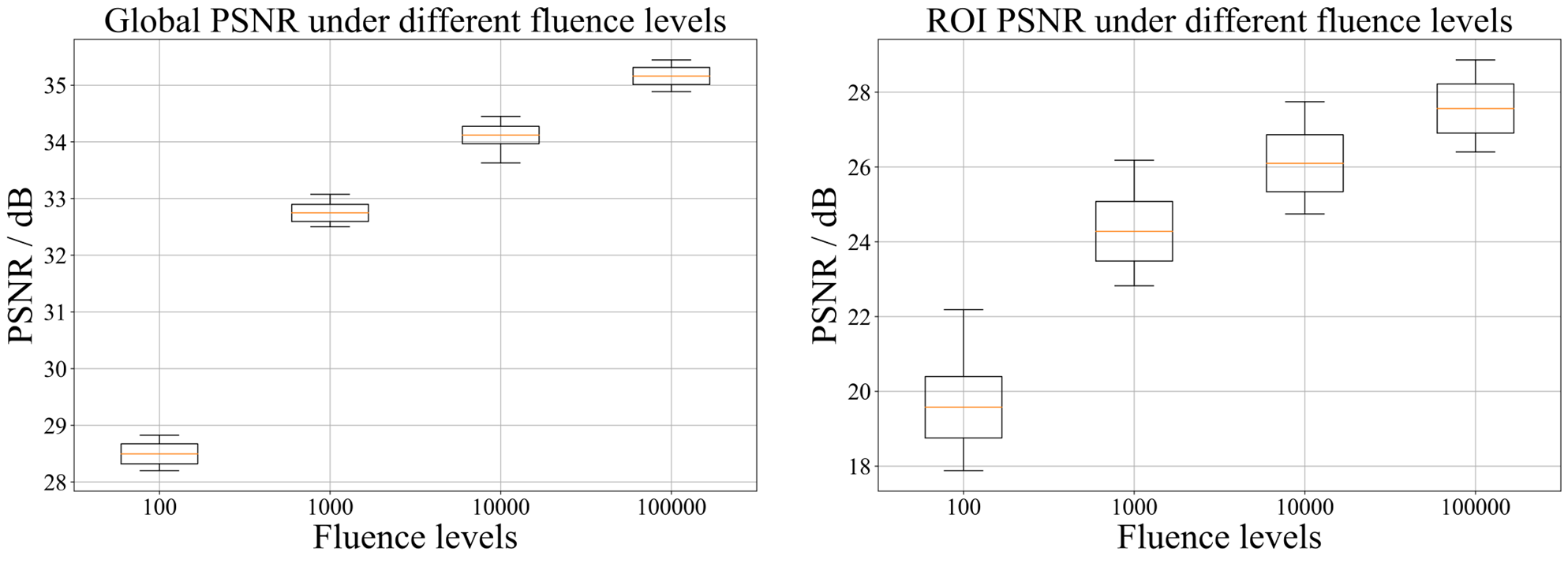}
   \end{tabular}
   \end{center}
   \caption[Multiple noise levels Box Plot]
   {\label{fig:multiple_box_plot}{ Boxplot of global and ROI PSNR for $400$ samples generated under each fluence level. The orange middle line represents the mean PSNR. The box shows the $\pm~1.5$ standard deviation, and the top and bottom whiskers show the maximum and minimum PSNR.}}
\end{minipage}
\end{figure}

{The distributions of global and ROI PSNR results of $400$ samples under each fluence level are shown in Figure \ref{fig:multiple_box_plot}. The global and ROI PSNR trends are similar with better performance at higher noise levels. Because errors are more pronounced at edges and in high spatial frequency textures, the lung ROI PSNR values are lower than the global values. The distribution of performance across DPS samples is illustrated with $\pm~1.5$ standard deviation box, and the maximum and minimum performance is shown with the brackets/whiskers. Interestingly, the level of variation in PSNR is very similar across fluence levels. We see slightly more variability in performance across samples when looking at the lung ROI at the lowest fluence (e.g. the standard deviation is slightly larger). Similarly, the lowest $I_0$ shows some asymmetry in the local performance with a longer ``tail'' in the distribution for higher PSNR - suggesting a slight skew in the distribution toward higher performance.}

\begin{figure}[h]
\centering
\begin{minipage}{\textwidth}
   \begin{center}
   \begin{tabular}{c} 
   \includegraphics[height=8.2cm]{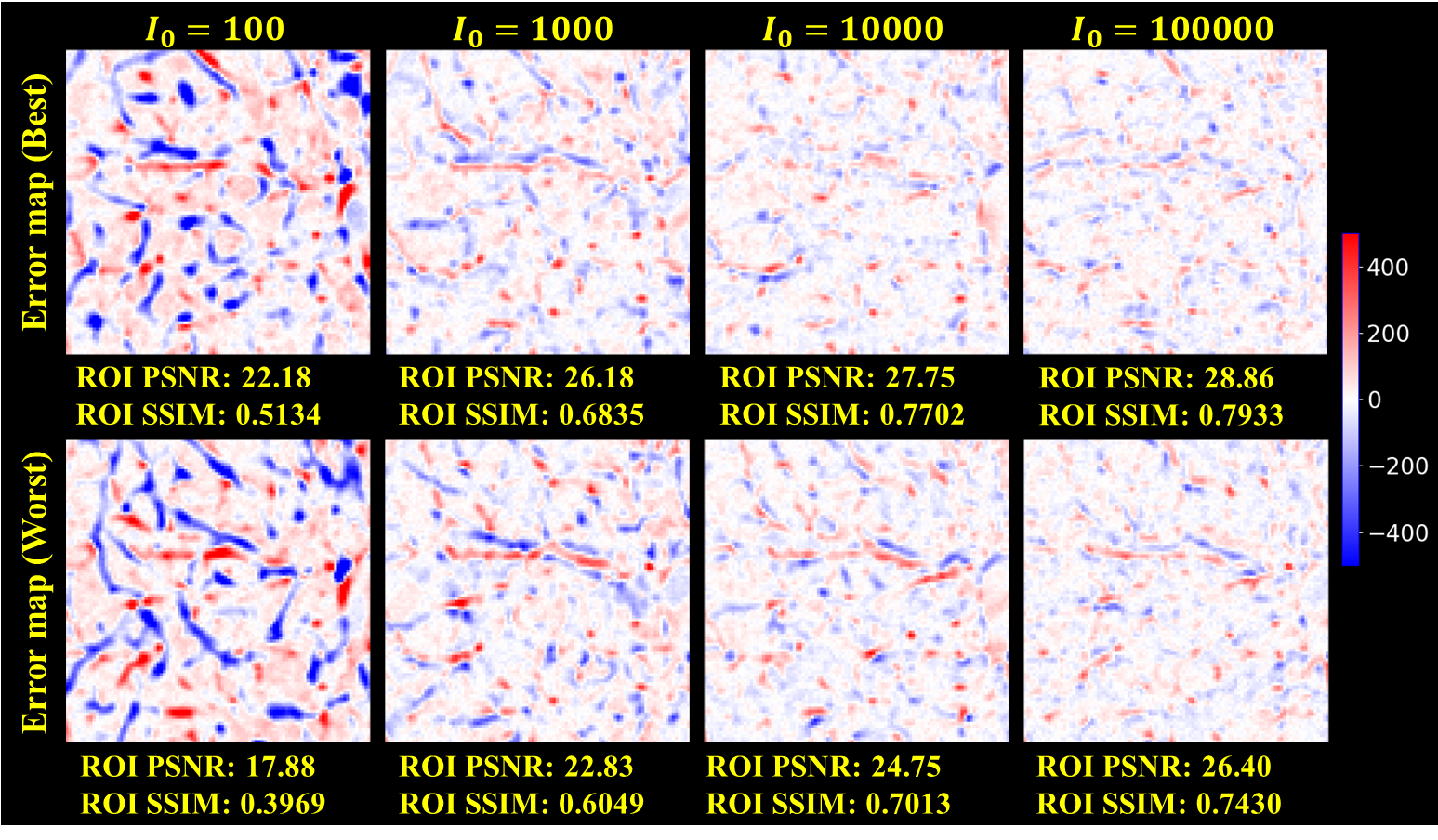}
   \end{tabular}
   \end{center}
   \caption[Multiple noise levels Best Worst]
   {\label{fig:multiple_best_worst}{ 
   Error maps for the best (top row) and worst (bottom row) case DPS samples in terms of ROI PSNR under different fluence levels. Colormap: W/L = 1000/0 HU for all images.}}
\end{minipage}
\end{figure}

{To visualize maximum differences in performance, Figure~\ref{fig:multiple_best_worst} shows the error maps for the best and worst ROI PSNR for each fluence levels. The difference between best and worst is most pronounced for the lowest fluence level; however, significant errors are still present even in the best case. There are common errors across both fluence levels and within the sample distributions (notably for some horizontal features near the middle of the ROI). At higher fluence levels, errors appear to be more consistent between the best and worst case outputs.}

\subsection{OS-DPS Nonlinear Evaluations}
\begin{figure}[h]
   \begin{center}
   \includegraphics[height=0.35\textwidth]{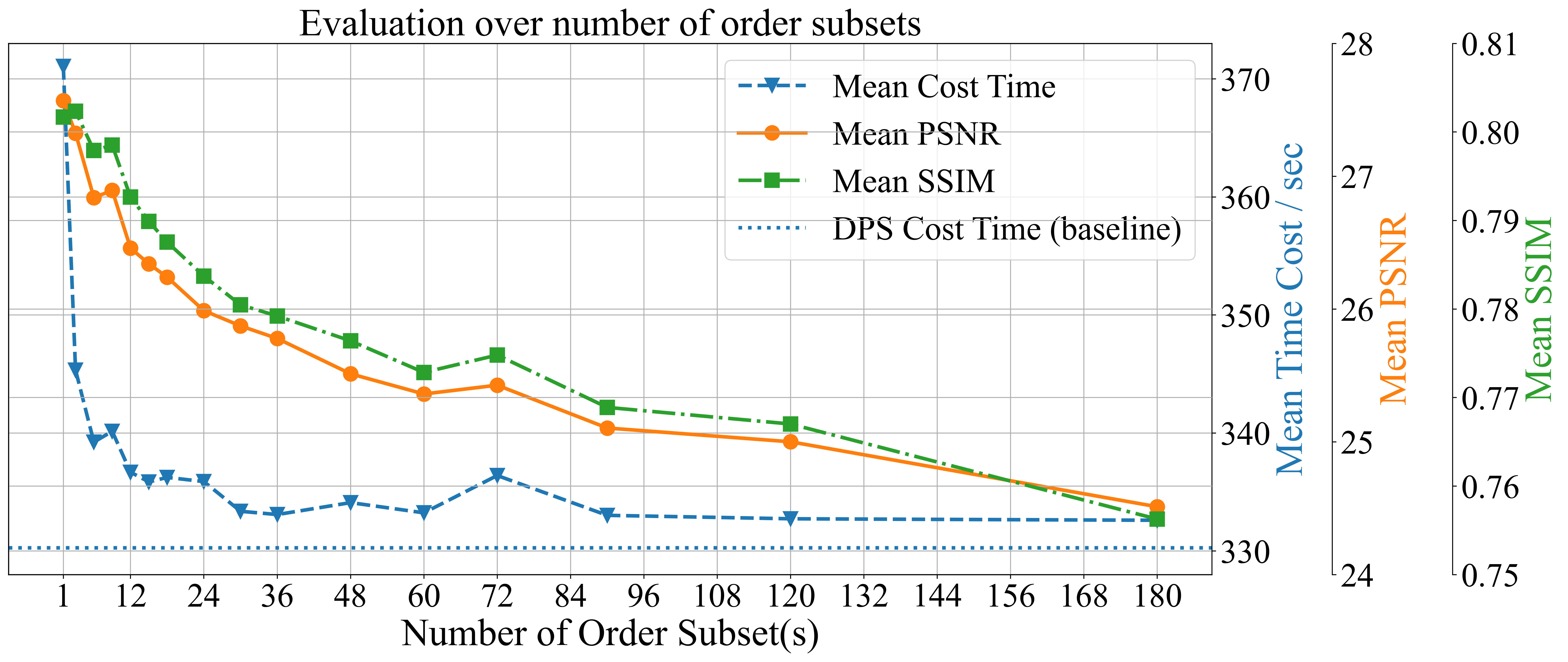}
   \end{center}
   \caption[OS DPS plot] 
   { \label{fig:os_plot} 
The mean computation time, SSIM, and PSNR over the 16 samples generated under different number of order subsets with $1000$ iterations on DDPM sampler. Mean cost time is colored as blue dashed line with triangle markers, mean PSNR is colored as orange solid line with circle markers, mean SSIM is colored as green dot dashed line with square markers, time cost by the unconditioned DPS is colored as blue dotted line.}
\end{figure}

\noindent Figure \ref{fig:os_plot} illustrates the mean {cost} time, SSIM, and PSNR of {16} samples for OS-DPS Nonlinear with different numbers of ordered subsets. Note that the computation time reached a plateau when the OS = {36} at around {$333.09$} seconds with a peak at OS=1 at {$371.05$} seconds. This represents a speed-up of around {$10.24\%$}. {The general runtime of unconditioned DPS is also drawn on the plot with blue dotted line, which takes $330.28$ seconds.} Recall that the ordered subsets approach is only speeding up the construction of likelihood term in Equation (\ref{eq:reverse_sampling}), indicating that the reverse diffusion update in the DPS Nonlinear reverse process remain a significant contributor to overall computation time. We did not observe a significant drop on the SSIM and PSNR as the number of subsets increased. The mean SSIM drop {$\sim 5.6\%$ from $0.8017$ to $0.7563$} and mean PSNR drop {$\sim 11.1\%$ from $27.56$ dB to $24.51$. Note that OS-DPS Nonlinear was demonstrated using LEAP version 0.8, while it cost around the same time (from $347.5$ seconds for OS=1 to $335.4$ for OS=180) for different ordered subsets under this low-dose scenario setting on LEAP version 1.0 and above. This speed-up was attribute to the CUDA optimization since LEAP version 1.0 where adaptive kernel was applied in single slice projection/backprojection, making the OS-DPS less sensitive to ordered subsets under this single projection setting. Nevertheless, the impact of ordered subsets on DPS will presumably be higher for larger data and volumes, especially fully 3D work in the future.} 

\begin{figure}[h]
   \begin{center}
   \begin{tabular}{c} 
   \includegraphics[height=10.8cm]{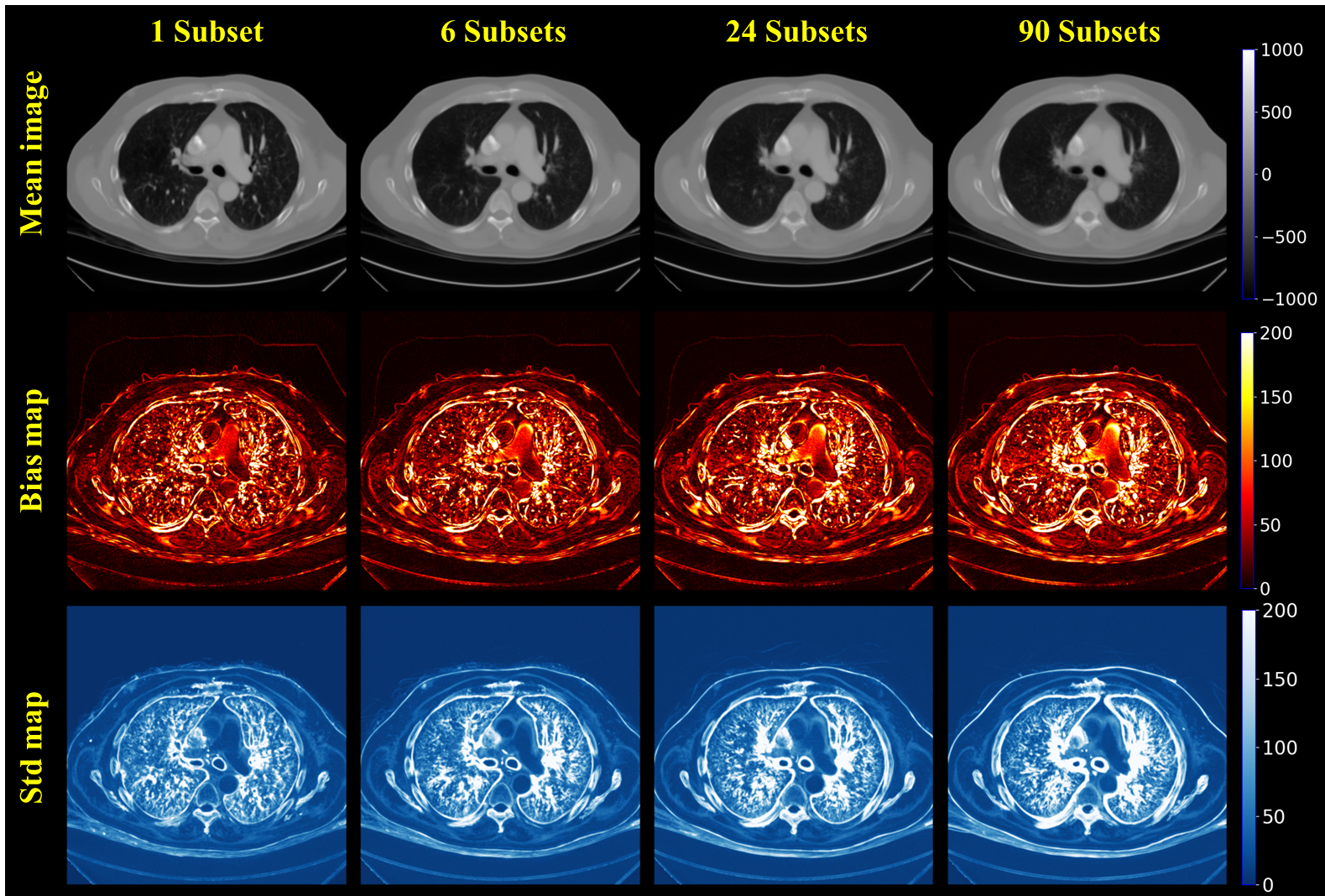}
   \end{tabular}
   \end{center}
   \caption[OS DPS results] 
   { \label{fig:OS_sample_images} 
Comparison of sample mean image, bias image, and mean variance image under different order subsets. The first row of images are the sample mean image of different order subsets, the second row of images are the bias image under different order subsets, and the third row of images are the mean variance image of different order subsets. Colormap: W/L = 2000/0 HU for mean and bias image, W/L = 1000/0 HU for bias image, and W/L = 500/250 HU for standard deviation image.}
   \end{figure} 

To visualize OS-DPS reconstruction results, we chose to plot the mean, bias, and standard deviation images for a sampling of different numbers of ordered subsets ({equal to 1, 6, 24, and 90}). These results are shown in Figure \ref{fig:OS_sample_images}. Again, the highest variability across all approaches lies in small lung features and at tissue boundaries. In general there appears to be an increase both in variability and bias as larger number of subsets are used.

{Table \ref{tab:runtime_compare} shows a runtime comparison using a single NVIDIA GeForce RTX 3090Ti GPU for image reconstruction. A comparison of various reconstruction approaches is presented, where we found that the proposed OS-DPS Nonlinear can perform a reconstruction in $333.09$ seconds, which is faster than DPS Nonlinear at $371.05$ seconds, DPS Linear at $370.12$ seconds, and DOLCE at $1199.80$ seconds. This illustrates that the ordered-subsets approach provides a modest improvement in runtime. Moreover, there does not appear to be a significant computation cost associated with the nonlinear model over the linearized approach. While we have not tried to optimize computation time for DOLCE, we find the DPS methods to be significantly faster based on our implementation of DOLCE.}

\begin{table}[ht]
\caption[Runtime Comparison]
{\label{tab:runtime_compare}
Runtime Comparison of single image reconstruction among different diffusion-based methods.}
\begin{center}       
\begin{tabular}{cc} 
\hline
\rule[-1ex]{0pt}{3.5ex}  Methods & Runtime (sec)  \\
\hline
\rule[-1ex]{0pt}{3.5ex}  OS-DPS Nonlinear (OS=6) & 333.09  \\
\rule[-1ex]{0pt}{3.5ex}  DPS Nonlinear (OS=1) & 371.05   \\
\rule[-1ex]{0pt}{3.5ex}  DPS Linear & 370.12   \\
\rule[-1ex]{0pt}{3.5ex}  DOLCE & 1199.80  \\
\hline
\end{tabular}
\end{center}
\end{table}

\section{Discussion and Conclusion}

We have demonstrated a new application of diffusion posterior sampling where a nonlinear measurement model is specified to combine prior knowledge of CT imagery from unsupervised training with a reverse process that is conditioned on the data likelihood model. We demonstrated the technique on {multiple systems to illustrate both the general applicability to different measurement models and noise levels, but also to show the advantages over current diffusion-based methods which rely on simplified (linearized) measurement models and/or conditional training of the diffusion network.} 

In an investigation of the ordered-subsets variant of DPS Nonlinear, we found that OS-DPS Nonlinear does permit modest gains in terms of computational speedup and/or image quality (in a scenario of limited reconstruction time). We note that our implementation of DPS Nonlinear was careful to use projectors and backprojectors that are PyTorch compliant to reduce inefficiencies associated with GPU-CPU transfers or other data formatting issues. While these computational advantages may be enhanced with a more thorough optimization of all algorithm parameters and may be more significant for larger datasets (e.g. 3D reconstruction where the projection operations are more costly); these results also suggest the need for acceleration of the computations associated with the reverse diffusion process. 

{Like other generative models, the proposed DPS Nonlinear technique is subject to hallucinations. We have illustrated such errors in an brief analysis of the distribution of DPS samples and found that the potential for these kinds of misrepresentation in the image increase in frequency and size as data fidelity/fluence decreases for DPS Nonlinear. While DPS grounds image reconstruction using a physics-based measurement model and provides a potential way to investigate hallucinations through multiple samples, users of such generative techniques should be aware of the potential for hallucination. How to address, analyze, and present data from generative methods remains an extremely important ongoing research area.} 


While the preliminary data shown here illustrates the potential of the DPS approach, the prospects for DPS Nonlinear application across imaging systems is very broad.  {We continue to investigate applications where the nonlinear model is a critical element to reconstruction. Examples include CT systems with projection blur from the detector\cite{li2024CT}, x-ray source, and/or gantry motion; as well as spectral CT\cite{xiao2024CT} that combines reconstruction and material decomposition. Such scenarios can be difficult (or impossible) to linearize - making the DPS Nonlinear approach an attractive option for reconstruction.}

\section* {Disclosures} 
The authors have no relevant financial interests or conflicts of interest to disclose.

\section* {Code, Data, and Materials Availability} 
Research code is currently not available. Research data is generated based on {CT Lymph Nodes Dataset \cite{roth2014new}}.

\section* {Acknowledgments}
This work was supported, in part, by NIH grant R01CA249538. Thanks to Xiaoxuan Zhang (University of Pennsylvania, Philadelphia, PA, USA), Peiqing Teng (Johns Hopkins University, Baltimore, MD, USA).


\bibliography{report}   
\bibliographystyle{spiejour}   




\vspace{2ex}\noindent\textbf{Shudong Li} is a graduate student currently at Tsinghua University. He received his BS degree in electronic Engineering from China Agricultural University in 2022. He received MS degree in biomedical engineering from Johns Hopkins University in 2023. His current research interests include deep learning and machine learning in Medical Imaging.

\vspace{2ex}\noindent\textbf{Xiao Jiang} is a PhD student currently at Johns Hopkins University. He received his BS degree and MS degree in Physics from University of Science and Technology of China in 2019 and 2022. His research focus on advanced processing methods for spectral imaging in x-ray imaging applications.

\vspace{2ex}\noindent\textbf{Matthew Tivnan} is an investigator at Harvard Medical School / Massachusetts General Hospital. He received his BS degree in electronic engineering and physics from Northeastern University in 2017, and his PhD degree in Johns Hopkins University in 2023. His current research interests involve the development of advanced instrumentation and reconstruction algorithms for Spectral CT.

\vspace{2ex}\noindent\textbf{Grace J. Gang} is an Assistant Professor in the Radiology Department of the University of Pennsylvania. She maintains an Adjunct Professor position in the Biomedical Engineering Department at Johns Hopkins. She leads research in collaboration with the AIAI Lab in task-based optimization to help design novel acquisition approaches and reconstruction methods for CT and CBCT. Her research interests also include the development and analysis of advanced reconstruction methods that integrate prior image information.

\vspace{2ex}\noindent\textbf{Yuan Shen} is a full professor in the Electronic Engineering Department at Tsinghua University. He received the BS degree in EE from Tsinghua University, and the MS and PhD degrees in EECS from the Massachusetts Institute of Technology. His current research focuses on network localization and navigation, integrated sensing and control, multi-agent systems, and bioinformatics. He was the elected Chair (2019–2020) for the IEEE ComSoc Radio Communications Technical Committee, and serves as Editor for several IEEE journals and symposium TPC Co-Chair for a number of IEEE conferences.

\vspace{2ex}\noindent\textbf{J. Webster Stayman} is an associate professor in the Biomedical Engineering Department at Johns Hopkins University. He has led efforts on the development of hardware and software for various imaging systems, including dedicated and interventional cone-beam-computed tomography devices. His research interests include system analysis and optimization, task-driven performance assessments, advanced algorithm development and machine learning, device design and prototyping, and clinical translation.

\vspace{1ex}
\noindent Biographies and photographs of the other authors are not available.

\listoffigures
\listoftables

\end{spacing}
\end{document}